\documentclass[12pt,a4paper,final]{iopart}

\usepackage{iopams}
\usepackage{graphicx}
\DeclareGraphicsExtensions{.eps}
\usepackage[breaklinks=true,colorlinks=true,linkcolor=blue,urlcolor=blue,citecolor=blue]{hyperref}
\begin{document}
\title[Nanotechnology (2015)]{Modulating spin relaxation in nanowires with infrared light
at room temperature}

\author{Md. Iftekhar Hossain$^{1}$, Saumil Bandyopadhyay$^{1,2}$, Jayasimha Atulasimha$^3$
and Supriyo Bandyopadhyay$^1$}
\address{$^1$Department of Electrical and Computer Engineering, Virginia Commonwealth University, Richmond, Virginia 23284, USA}
\address{$^1$Department of Electrical Engineering and Computer Science, Massachusetts Institute of Technology, Cambridge, Massachusetts 02139, USA}
\address{$^2$Department of Mechanical and Nuclear Engineering, Virginia Commonwealth University, Richmond, Virginia 23284, USA}
\eads{\mailto{sbandy@vcu.edu}}

\begin{abstract}
Spintronic devices usually rely on long spin relaxation times 
and/or lengths for 
optimum performance. Therefore, the ability to modulate these quantities with an 
external agent offers unique possibilities. The dominant 
spin relaxation mechanism in most technologically important semiconductors 
is the D'yakonov-Perel' (DP) mechanism which vanishes if the spin carriers 
(electrons) are confined to a single conduction subband in a quantum wire
grown in certain crystallographic directions, or polycrystalline quantum wires. Here, we report modulating 
the DP spin relaxation rate (and hence the  spin relaxation length)
in self assembled 50-nm diameter 
InSb nanowires with infrared light at room temperature. 
In the dark, almost all the electrons in
the nanowires are in the lowest conduction subband at room temperature, resulting in 
near-complete absence of DP relaxation. This allows observation of spin-sensitive effects in the 
magnetoresistance. Under infrared illumination, electrons are photoexcited to 
higher subbands 
and the DP spin relaxation
mechanism is revived, leading to a three-fold decrease in the spin relaxation length. Consequently, the
spin sensitive effects are no longer observable under illumination. This phenomenon may have applications
in spintronic 
room-temperature infrared photodetection. 
\end{abstract}

\maketitle

\section*{Introduction}

It is well-known that the D'yakonov-Perel' (DP) spin relaxation mechanism  \cite{dyakonov} is completely absent in a
semiconductor quantum wire of certain crystallographic orientations if
 only the lowest conduction subband state is occupied by electrons (spin carriers) 
\cite{pramanik, book}. In narrow gap semiconductors (e.g. InSb, InAs), or in polycrystalline samples, where the Rashba 
interaction 
\cite{rashba} would be far stronger than the Dresselhaus interaction \cite{dresselhaus}, single subband occupancy will
nearly eliminate any DP relaxation regardless of the wire's crystallographic orientation.
Since the DP relaxation is the dominant spin relaxation mechanism in compound semiconductors 
like InSb \cite{dyakonov2, averkiev, averkiev2, kainz, song, patibandla} at room temperature, 
its elimination can 
increase the spin relaxation time and spin relaxation length in these semiconductors considerably. That is an important 
result since spin-relaxation is the spoiler in most spin-based devices and applications.

There have been theoretical predictions  that the DP spin relaxation rate will decrease in quantum wires 
compared to bulk or quantum well systems \cite{bournel,malshukov,kim,pareek} and this has been borne out by experiments \cite{saumil, holleitner}. 
Ref. \cite{saumil} studied InSb nanowires where $\sim$96\% of the electrons were expected to reside in the 
lowest conduction subband  and found that at room temperature
the total spin relaxation time had increased
by an order of magnitude over that reported in bulk or quantum wells, despite poorer electron mobility in the nanowires.
This result was attributed to strong suppression of the DP relaxation. However, these experiments are
not confirmatory proofs that the elimination/suppression was purely due to single/few subband occupancy since confirmation
would require showing that the spin relaxation time is {\it controllably} increased by {\it controllably}
exciting electrons from the lower to the higher subbands.
 
In this paper, we report controlled modulation of the DP 
relaxation rate at room temperature in 50-nm diameter InSb nanowires by varying 
the subband population with an external agent. This leads to controlled modulation of the 
net spin diffusion length. The obvious 
approach to  vary subband population controllably would have been to use split-gate 
quantum point contacts that allow varying the nanowire width (and hence the subband population) with a gate potential 
\cite{dutch, english}. In comparison, our approach
 to use infrared (IR) light to excite electrons 
continuously 
to higher subbands from the lowest one -- thereby causing multi-subband transport -- is much easier since it does not require 
making a gate contact. 
Furthermore, varying the nanowire width with a split gate, or simply 
using different nanowires with different diameters \cite{holleitner}, does not provide an unambiguous picture of how the
DP relaxation rate
depends on subband occupation because changing diameter could affect the carrier mobility (which is mostly
governed by interface roughness scattering) and thus affect other spin relaxation rates, notably the Elliott-Yafet \cite{elliott}. 
This problem is somewhat mitigated if the subband occupation is changed with light.
The 
k-selection
rule guarantees that the electron's momentum before and after photon absorption are approximately 
equal,
 so light does not affect the Elliott-Yafet spin relaxation directly because
 that requires a 
change in electron momentum. However, occupation of higher subbands can change carrier mobility by promoting 
inter-subband scattering and this can affect the Elliott-Yafet rate. Fortunately, this effect is weak because the mobility 
in the nanowires we fabricate is governed primarily by interface roughness scattering and not inter-subband scattering. Therefore,
any light-induced change in mobility will be small.
Light may also induce transitions between orthogonal spin states in the same or different subbands
when spin-orbit interaction is present
\cite{upadhyaya} and thus cause some spin relaxation, but the matrix elements for such transitions are 
so weak that this effect can be ignored. Therefore, the primary effect of light on spin relaxation 
in single-subband quantum wires is to change the 
DP relaxation rate by causing a transition from single-subband transport to multi-subband transport.
If this change turns out to be significant, then this effect can be used to implement a
room-temperature IR photodetector 
in the following way:

\begin{figure}
\includegraphics[width=3.4in]{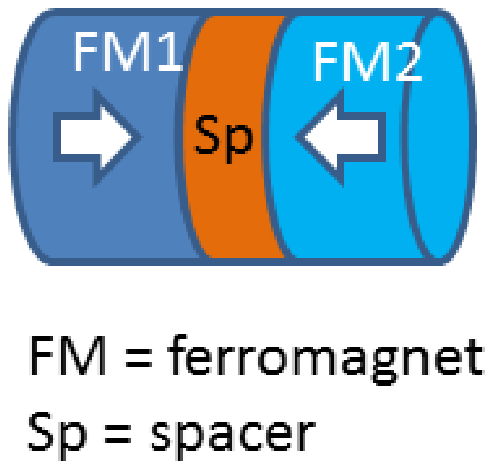}
 \caption{A nanowire spin valve whose two ferromagnetic contacts have opposite signs of tunneling spin polarization.}
\end{figure}
Consider a tri-layered nanowire spin-valve as shown in Fig. 1 whose two ferromagnetic 
contacts have opposite signs of tunneling spin-polarization.
They could be made of Co and Fe. Applying a strong magnetic field along the nanowire
axis magnetizes both contacts in the direction of the field. Spins parallel to the magnetic field will become 
majority spins in the contacts while those anti-parallel will become minority spins. Assume that the 
left contact 
is Co and the right contact is Fe. Under a suitable bias, the left contact preferentially injects its 
minority spins into the middle spacer layer via tunneling through the Schottky barrier 
at the Co/spacer interface since Co has a negative tunneling spin polarization at the Fermi energy 
\cite{tsymbal}.
These spins 
cannot transmit easily through the Schottky barrier at the right contact interface into the Fe contact
since they are minority spins and Fe has a positive tunneling spin polarization \cite{tsymbal}. 
However, if the injected spins 
flip (relax) in transit, then they will become majority spins at the right contact and 
transmit easily \cite{hall}. If the 
spin relaxation
rate can be increased with IR light, then that will
increase the likelihood of the injected spins flipping and thereafter being 
transmitted easily by the right contact. This will cause an increase in the
current through the nanowire under IR illumination, which is the physical basis of the 
photodetection mechanism. We can estimate the 
light-to-dark current contrast ratio of the photodetector as follows:

Assume that IR light induces only intra-band transitions in the conduction band of the spin-valve's spacer layer 
and not inter-band 
transitions from the valence to the conduction 
band which would have changed the carrier concentration. This assumption
 would be particularly
true for wide-gap semiconductor spacers where the bandgap vastly exceeds the IR photon energy. 
Also assume that the current in the spacer is mostly due to drift and not 
diffusion. In that case, it is given by

\begin{equation}
I = {{e v_d n_l}\over{4}} \left [ 1 + \zeta_1 \zeta_2 - 2 \zeta_1 \zeta_2 e^{- {{L}\over{L_s}} } \right ],
\end{equation}
where $e$ is the electron charge, $v_d$ is the drift velocity of electrons,
$n_l$ is the linear electron concentration in 
the spacer layer,
$L$ is the length of the spacer layer, $\zeta_1$ is the spin injection
efficiency at the injecting (left) 
contact, $\zeta_2$ is the spin detection efficiency at the detecting 
(right) contact, and $L_s$ is the spin relaxation length 
in the spacer layer, ensemble averaged over electron velocity.

In deriving the above relation, we made two assumptions. First, even in single subband
transport (when DP relaxation is absent), spin-orbit interaction will make the spins precess as they travel,
but the angle $\theta$ by which they precess in traversing the distance $L$ is the same 
for all electrons (coherent rotation) \cite{book}. We neglected this precession
since $\theta$ is very small when $L$ is small. Second, in multi-subband transport when DP relaxation 
is present, different 
electrons will precess by different amounts in traversing the distance $L$ (incoherent rotation) which 
causes spin relaxation in space and is the basis of 
DP relaxation. 
We assume  that this relaxation 
is captured by an exponential decay of spin polarization with distance. This, however, may be only approximately true \cite{sandipan_prb}.

Assuming that IR light does not change $\zeta_1$ or $\zeta_2$ appreciably, the light-to-dark
contrast 
ratio will be

\begin{equation}
{\tt Contrast~ratio} = {{I_{light}}\over{I_{dark}}} = {{1 + \zeta_1 \zeta_2 - 2 \zeta_1 \zeta_2 
e^{- {{L}\over{L_l}} }
}\over{1 + \zeta_1 \zeta_2 - 2 \zeta_1 \zeta_2 e^{- {{L}\over{L_d}} }}} ,
\label{ratio}
\end{equation}
where $L_l$ and $L_d$ are the spin relaxation lengths under illumination and in the dark, respectively. 
In the event
$L_l, L_d \gg L$, the above expression simplifies to 
\begin{eqnarray}
{\tt Contrast~ratio} & \approx &  {{1 - \zeta_1 \zeta_2 + 2\zeta_1 \zeta_2 \left (L/L_l \right )}\over{1 - \zeta_1 \zeta_2 + 2\zeta_1 \zeta_2\left (L/L_d \right )}} \nonumber \\
& \approx & {{L_d}\over{L_l}}
\left ( {\tt if}~ \zeta_1 \approx \zeta_2 \approx 1 \right ).
\label{ratio1}
\end{eqnarray}

\section*{Experimental details}

In order to demonstrate modulation of the DP spin relaxation rate with IR light and additionally to lay the foundation for a
spintronic 
IR photodetector, we fabricated arrays of Co-InSb-Ni nanowires of $\sim$50 nm diameter.
First, nanoporous anodic alumina 
films, containing {\it parallel} arrays of 50-nm diameter cylindrical nanopores, were produced by
anodizing 99.999\% pure aluminum foils in 0.3M 
oxalic 
acid at room temperature. Prior to 
anodization, the foils were 
electropolished in a solution of perchloric acid, butyl
cellusolve, ethanol and distilled water 
to reduce the 
surface roughness to $\sim$3 nm \cite{bandy}.
The anodization was carried out at 40 V dc for 15 minutes; 
the voltage was
 then gradually reduced to 15 V at the rate of
 0.1 V/sec and held for 10 minutes before terminating the 
anodization abruptly.
This step anodization process \cite{furneaux} allows us to remove the thin 
alumina  layer that forms 
at the bottom 
of the pores 
(which is a barrier to dc current flow) by later soaking the
porous film in 5\% phosphoric acid 
for 30 minutes. 

After removal of the barrier layer,
cobalt is dc-electrodeposited selectively within the pores from a 
solution of 28.09 gm of
 CoSO$_4$$\cdot$7H$_2$O and 7 gm of boric 
acid dissolved in 1 liter of distilled water. 
Next, InSb is 
dc-electrodeposited from a solution of 0.15M 
InSO$_4$, 0.1M SbCl$_3$, 0.17M Na$_3$C$_6$H$_5$O$_7$ and 
0.36M C$_6$H$_8$O$_7$ (citric
acid) dissolved in 250 ml of distilled water \cite{insb}. 
Finally, Ni is dc-electrodeposited 
from 
a solution of 26.27 gm of 
NiSO$_4$$\cdot$6H$_2$O and 7 gm of boric acid dissolved in 1 liter of distilled water. The Co 
and 
InSb depositions
are carried out at 3 V dc for 30 seconds and 1 minute, respectively, while the Ni deposition is 
carried out at 5 V dc
for 4 minutes to fill the pores to the brim or slightly overfill and make Ni spill out on the 
surface. 
A 50-nm thick layer of Ni is then electron-beam-evaporated on the surface through  a mask.
The nanowires are 
electrically contacted from the top and bottom with copper 
wires attached to the top surface (evaporated Ni) and the 
bottom aluminum substrate
with silver paste. Since the 
nanowire density is $\sim$10$^{10}$ cm$^{-2}$ and the contact area 
is a circle of roughly
6 mm diameter, about 3$\times$10$^9$ wires are 
contacted in parallel and probed in electrical 
measurements.

Fig. 2 shows transmission electron micrographs (TEM) 
of isolated nanowire spin valves. The TEM
samples were 
prepared by dissolving out the porous alumina matrix hosting the nanowires in dilute NaOH and then  capturing 
the released wires on TEM grids by soaking the grids in the 
solution. Since Co and Ni both have atomic densities of 
roughly 9.1$\times$10$^{22}$ cm$^{-3}$, while
InSb has an atomic density of 2.94$\times$10$^{22}$ cm$^{-3}$, the InSb 
region appears more transparent than 
the Co or Ni regions in a bright field TEM image.

\begin{figure}
\includegraphics[width=5.1in]{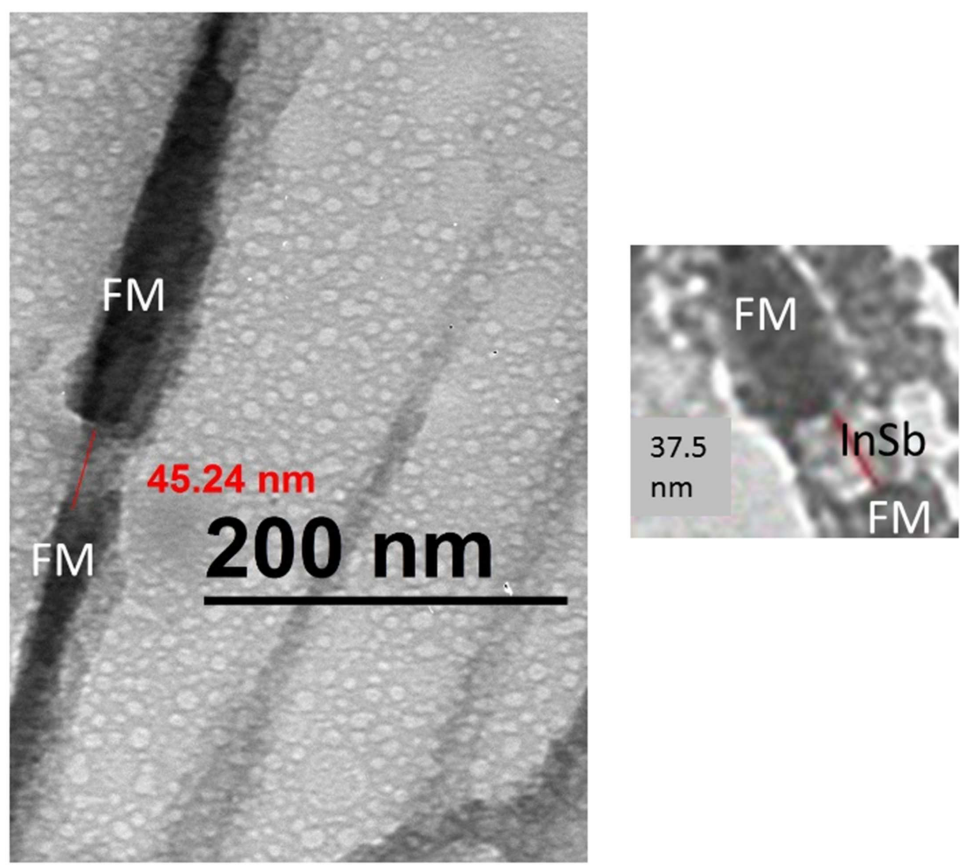}
\caption{Bright field transmission electron micrograph of Co-InSb-Ni nanowire spin valves formed within 
anodic alumina 
pores of 50 nm diameter (FM=ferromagnet). The InSb spacer layer length varies from wire to wire because of the 
fabrication process, 
but the spread is not large and the average spacer layer length is $\sim$40 nm. In the TEM
samples, 
the Co electrodeposition time was intentionally increased to 4 minutes in order to obtain a long 
Co section for sufficient contrast
that will allow unambiguous determination of the InSb spacer layer's length. 
In the actual spin valves, the Co section is much shorter because the electrodeposition
time was only 30 seconds.}

\end{figure}

Energy dispersive x-ray spectroscopy (EDS) of the nanowire spin valves nestled within the alumina matrix
 is 
presented in the 
supplementary material.
We found clear peaks due to Ni, Co, In and Sb (associated with the nanowire spin 
valves), as well as Al and O peaks 
caused by the alumina matrix and the 
aluminum substrate. The magnetization curves for porous alumina samples filled 
with just Co (30 seconds electrodeposition) and just Ni (4 minutes electrodeposition) 
are shown in the supplementary material and confirm that 
the nanocontacts are ferromagnetic and have non-zero remanence,
albeit with small coercivity. The single subband occupancy of the nanowires (in the dark) was established in Ref. 
\cite{saumil}.



\section*{Results and Discussion}

 Since the Ni electrodeposition is carried out for a relatively long duration, the Ni nanocontacts have 
cylindrical shapes conforming to the pores in which they are produced and their axes are collinear with the nanowires' common axis. 
Therefore, their easy axes of magnetization is always along the wire axis. However,
since the Co deposition is  
carried out for only 30 seconds, the Co
layers in the spin valves are very short and therefore do not necessarily have cylindrical shapes with
their easy axes of magnetization along the nanowire axis.
 Consequently, their magnetizations may not be
collinear with the nanowires' common axis when no magnetic field is present.  At a high enough magnetic field directed along the wire axis, the magnetizations of the short Co
contacts will ultimately rotate to align 
along the field and hence along the nanowire axis. 
For each nanocontact, this will happen abruptly when the magnetostatic energy due to the magnetic field just overcomes its shape anisotropy energy barrier. 
If the shape anisotropy energy barrier does not vary too much from one Co nanocontact to another among the 3$\times$10$^9$ nanowires probed, then all nanocontacts undergo the rotation 
at nearly the {\it same} field. At that
threshold field, 
the majority spins in the Co and Ni contacts should suddenly become parallel, 
and the (spin-dependent) transmission 
probability of the electrons should jump, causing the resistance of the entire sample (consisting of 3$\times$10$^9$ nanowires in parallel) to 
drop {\it abruptly} at that field. 
On the other hand, if the shape anisotropy energy barriers vary significantly from one Co nanocontact to another, then different nanocontacts in a sample will
rotate at different fields and the resistance of the sample should drop {\it gradually} with increasing field and not abruptly.
However, in some circumstances, a very different behavior 
may be observed,
depending on the nature of the transport in the InSb layer. If there is a point defect site in
the InSb spacer layer close to one of the ferromagnetic electrodes and this defect has an energy level that is resonant with
the Fermi 
energy of that electrode, then an electron traversing the spacer layer will resonantly tunnel through this defect site and
that will effectively
{\it invert} the spin polarization of that ferromagnetic electrode \cite{tsymbal1}. 
In that case, the resistance of the spin valve will rise at the threshold field 
instead of dropping. 
This type of polarization inversion has been frequently
observed in ferromagnetic/paramagnetic 
nanojunctions of cross section smaller 
than 0.01 $\mu$m$^2$ grown 
by electrodeposition, as is the case here 
\cite{tsymbal1}. Therefore, if resonant tunneling occurs,
then we will expect a rise in the 
spin valve resistance (either abrupt or gradual) with increasing magnetic  field, whereas if no 
such tunneling occurs, then we will expect a drop. The sign of the magnetoresistance change 
therefore depends on the nature of transport. 
Both signs have been observed in different samples 
fabricated in the same run in the past because the impurity characteristics are beyond control \cite{pramanik-prb}. 
We too have observed both signs in our samples.

In Fig 3, we 
show the room-temperature magnetoresistance of one sample measured in the dark and under illumination 
by an infrared (IR) 
lamp radiating in the wavelength range 2-5 $\mu$m. 
The magnetoresistance was measured with the magnetic field directed parallel 
to the axes of the wires. The lamp was kept far enough 
away from the sample to avoid heating 
effects. The resistance was monitored as a function of time during illumination to see if it drifted with time. Since no
drift was observed, 
there are no discernible thermal effects on resistance due to the IR lamp.

\begin{figure}[!h]
\includegraphics[width=3.4in]{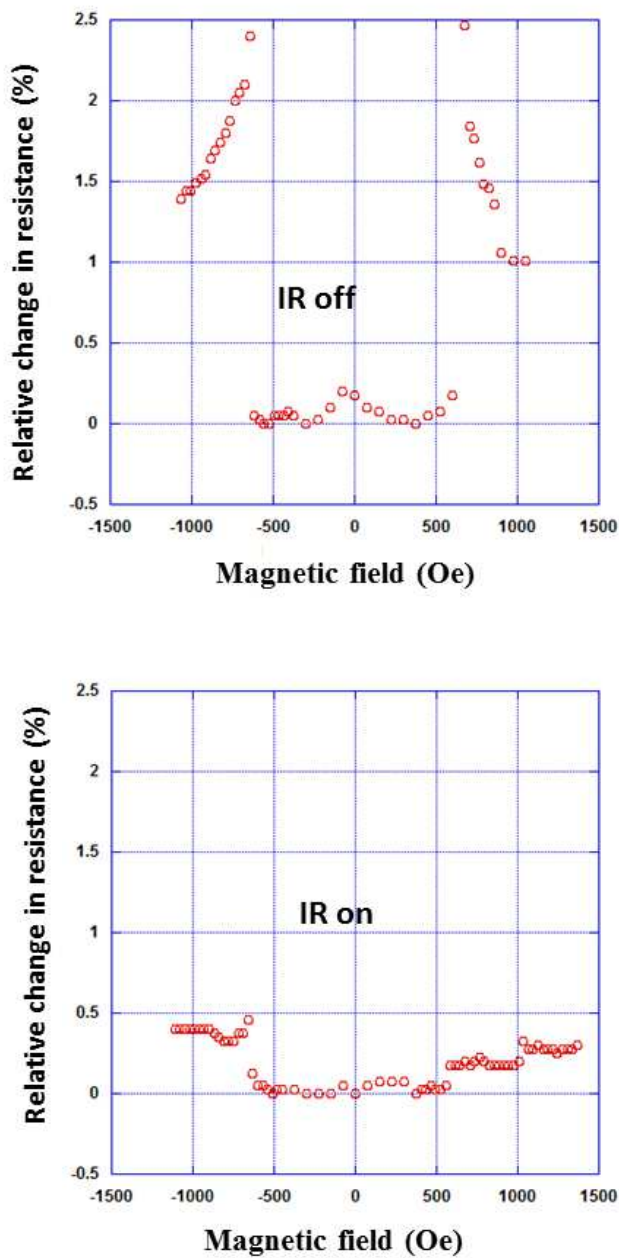}
\caption{Room-temperature magnetoresistance of a Co-InSb-Ni nanowire spin sample in the dark (above) and under illumination 
by an 
infrared lamp radiating in the wavelength range 2-5 $\mu$m (below). The zero-field dark resistance was 4.2 ohms.}
\end{figure}

\begin{figure}[!h]
\includegraphics[width=5.1in]{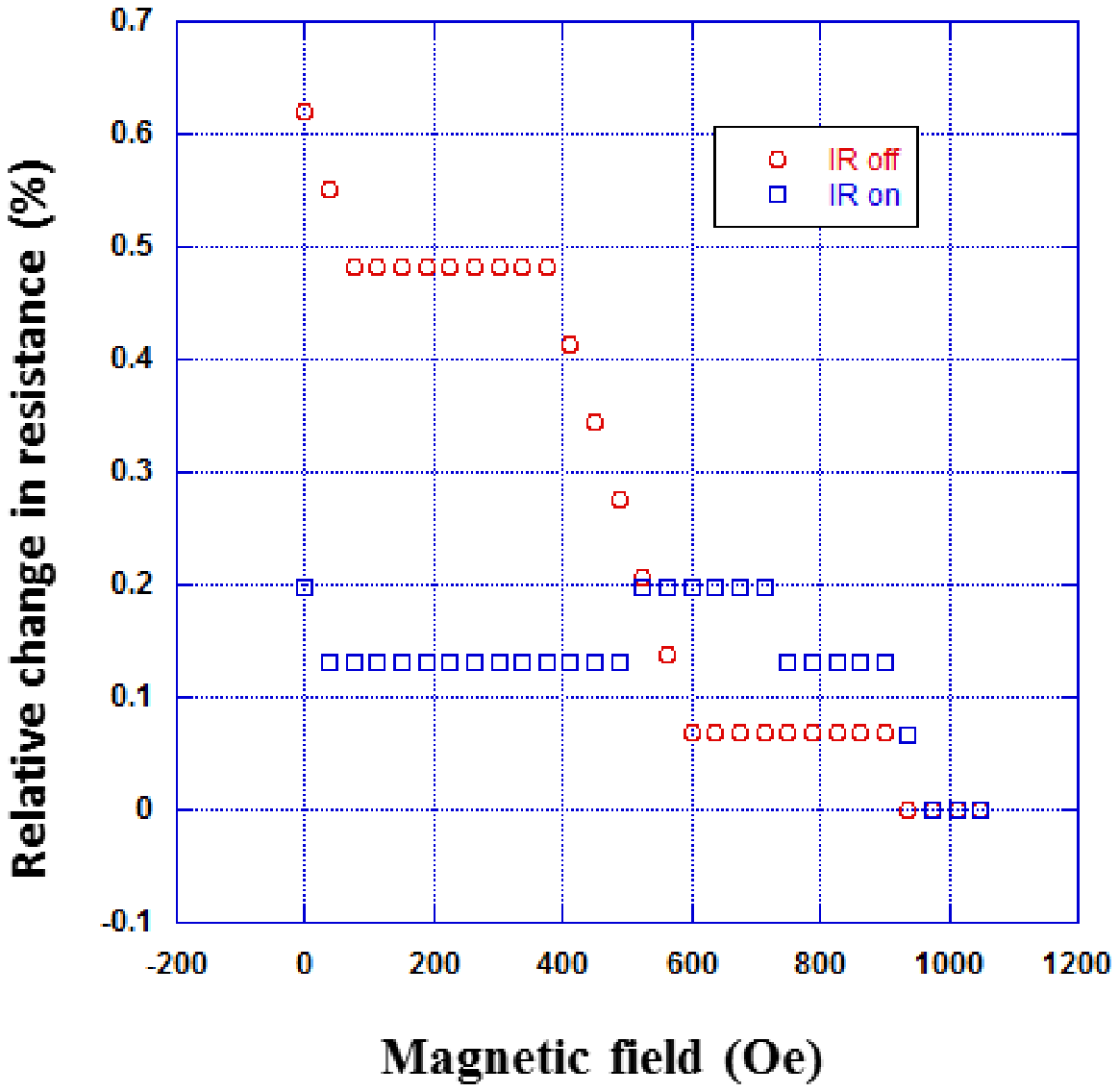}
\caption{Room-temperature magnetoresistance of another Co-InSb-Ni nanowire spin sample in the dark and under illumination 
by an infrared lamp radiating in the wavelength range 2-5 $\mu$m. The zero-field dark resistance was 9.7 ohms.}

\end{figure}

The data taken in the dark (top panel of Fig. 3)
show a reproducible abrupt 
jump in the resistance by 2.5\% when the magnetic field strength exceeds $\pm$650 Oe. Clearly, 
this is the 
threshold 
field at which the magnetizations of the Co nanocontacts overcome their shape anisotropy energy barriers and rotate 
to align along the nanowire axis. The fact that the resistance rise is abrupt indicates that the variation in the shape anisotropy 
energy barrier among the different Co nanocontacts in this sample is small. 

Note that beyond the threshold field, the resistance actually drops continuously.
This is probably due to the fact that the magnetic field 
increases the
energy separation between subbands and decreases inter-subband scattering in the InSb spacer layers,
thereby slightly decreasing the resistance. This is unrelated to any
spin-sensitive effect. However, the continuous 
decrease
 in magnetoresistance beyond the threshold field gives us additional confidence that the sudden step increase 
at the threshold field must be due to spin-polarized transport. 

When the magnetoresistance is
measured under constant IR illumination (bottom panel of Fig. 3), we observe the same resistance 
jump at around $\pm$650 Oe magnetic field, but 
this time the resistance change is only 0.4\%, which is considerably
less than that in the dark. The more than 6-fold decrease in the step can only happen
if spin-polarized transport has been weakened by IR light. The IR 
light causes multi-subband transport in the InSb spacer layers of the spin valves by exciting electrons 
to the higher subbands and therefore shortens the spin 
relaxation length by reviving DP relaxation.
The shorter spin relaxation length decreases spin polarization of the 
current and hence suppresses the step increase in the magnetoresistance at the threshold field.

We can obtain at least an order 
estimate for the average spin relaxation length in the InSb spacer layers -- both in the dark and under IR illumination -- from the modified Jullier\'e formula
for drift transport
\cite{julliere,vardeny}:

\begin{equation}
{{\Delta R}\over{R}} = - {{2 P_1 P_2 e^{-L/L_s}}\over{1-P_1 P_2 e^{-L/L_s}}} ,
\end{equation}
where $\Delta R/R$ is the relative change in resistance at the step, $P_1, P_2$ are the spin polarizations of the 
two contacts, $L$ is 
the average width of the InSb layer and $L_s$ is the average spin relaxation length. The 
TEM results in Fig. 2 show $L \simeq$40 nm. Because the 
Co 
nanocontacts are small, but the common Ni contact to the nanowires is large (6 mm diameter) and approximates bulk, we will assume $P_1$ = 0.1 (for Co)
 \cite{saumil,tinkham} 
and $P_2$ = 0.33 (for Ni) \cite{tsymbal}. Therefore, from the measured 
 $\Delta R/R$, we get $L_s$ $\approx$ 40 nm in this 
sample
 at room temperature in the dark. Under IR illumination, the quantity $\Delta R/R$ drops to 0.4\% from 
 2.5\%, indicating that the average spin relaxation 
length $L_s$ has decreased three-fold 
 to $\sim$14 nm. 
 Therefore, $L_d$ = 40 nm and $L_l$ = 14 nm. 

It is reasonable to assume that $P_1 \approx \zeta_1$ and $P_2 \approx \zeta_2$. If we use the values $\zeta_1$ = 0.1, $\zeta_2$ = 0.33, $L$ = 40 nm, $L_d$ = 40 nm and $L_l$ = 14 nm in Equation (2)
to find the relative change in resistance between dark and illuminated conditions, we find that the quantity
$\left (R_d - R_l \right )/R_l
= 2 \zeta_1 \zeta_2 \left ( e^{-L/L_d} - e^{-L/L_l} \right )/\left (1 + \zeta_1 \zeta_2 - 2 \zeta_1 \zeta_2 e^{-L/L_d} \right )$
= 2\%, where $R_d$ is the dark resistance and $R_l$ is the resistance under illumination. 
Experimentally, we measure this quantity to be also 2\% (see Fig 3), which indicates excellent 
agreement between theory and experiment.

In Fig. 4, we show the magnetoresistance of another sample 
measured in the dark and under IR illumination.
In the dark, this sample shows a more gradual resistance {\it drop} in the 
magnetic field range 400 Oe - 600
Oe -- more gradual probably because there is significant variation in the shape 
anisotropy energy barriers among the Co nanocontacts. The fact that the resistance decreases 
with increasing magnetic field
 indicates that the carriers in this sample 
transport without resonantly
tunneling through defect sites. The majority spins in the two magnetic nanocontacts begin to become parallel 
when the magnetic field exceeds 400 Oe, resulting in the gradual resistance drop. The drop is $\sim$0.4\%. 
Under 
illumination, there is no discernible change in the resistance with increasing magnetic field
(the random jumps of 0.1\% change show no systematic trend), 
indicating that no significant
spin-polarized transport is taking place since the average spin relaxation length has become much 
shorter
than the average spacer layer length. In this sample too, light had induced multi-subband transport, thereby reviving the DP relaxation and 
shortening 
the average spin relaxation length.
 Equation (4) yields that the spin relaxation length in this sample is 
$\sim$14 nm in the dark and immeasurably
small under illumination. These results are summarized in Table I.

\begin{table}[!t]
\caption{Spin relaxation lengths in the dark and under IR illumination}
\medskip

\begin{tabular}{ccc}
\hline
 
\hline
& $L_d$ (dark) & $L_l$ (illuminated) \\
 \hline
 
Sample 1 & 40 nm & 14 nm \\
 Sample 2 & 14 nm & Immeasurably small \\
 \hline
 
\hline
\end{tabular}
 
\end{table}

There have been previous reports of modulating the spin relaxation time and length in semiconductor quantum wells 
by modulating the Rashba spin-orbit interaction strength with an external electric field applied via a gate \cite{iba, wang}.
However, the mechanism there is different; spin relaxation time and length are not controlled by changing subband occupation, but by 
changing the strength of the relaxation source, namely the spin-orbit interaction strength. This may have transistor 
applications \cite{hall}, whereas the present effect has photodetector applications.

The IR source is a broadband lamp and radiates in the wavelength 
range between 2 and 5 $\mu$m corresponding to photon 
energies 
between 9.4 kT and 24 kT, while the effective bandgap of the InSb layer is $\sim$ 6.5 kT. 
Therefore, the IR illumination will induce both inter-band transition (valence-to-conduction band) and 
inter-subband 
transition 
within the conduction band of the InSb spacer layer. 
The resistance 
of the sample at zero magnetic field 
(when significant spin-polarized transport does not occur) however decreased 
by only 0.15\% under 
illumination indicating that
the inter-band transition (which will increase the electron and hole  population in the 
InSb spacer and therefore
decrease the resistance) 
is not significant. This is because the light intensity from
the IR lamp is 
too weak 
for significant interband transitions to occur. It is therefore also too weak to cause significant 
intraband (or inter-subband) transitions, 
and yet the spin relaxation length decreased by a factor of three under illumination.
This is due to the 
fact that even a slight departure from single subband transport can increase the DP relaxation rate considerably
and quench spin-polarized transport. It is an encouraging observation since it portends high detectivity and small noise 
equivalent power for 
photodetectors predicated on this effect.

\section*{Conclusion}

In this work, we have shown that it is possible to modulate the spin relaxation rate in a nanowire with light 
by controlling the D'yakonov-Perel' mechanism. The mechanism is suppressed in the dark owing to near single 
subband occupancy and revived under illumination as more subbands get populated by photoexcitation.

If the spin injection and detection efficiencies in these spin valves can be improved, and the Elliott-Yafet 
and other spin 
relaxation 
mechanisms suppressed by producing very high mobility samples with weak hyperfine interactions, 
then we can use 
this effect to implement a room-temperature IR detector with very high light-to-dark 
contrast ratio. Consider the situation 
when the spin injection/detection efficiencies approach 100\%.
From Equation (\ref{ratio1}), we get that the contrast ratio is 
$L_d/L_l \approx L_{DP}(s)/L_{DP}(m)
\rightarrow \infty$, where  $L_{DP}(s)$ is the DP relaxation length in 
single-subband transport (infinity) and 
$L_{DP}(m)$ is the DP relaxation length in multi-subband transport (finite). 
Such a photodetector will also 
ideally have nearly zero dark current and hence almost zero standby power dissipation, 
making it very attractive.
The experiments reported here lay the groundwork for
such a device.

\section*{Acknowledgement}

\noindent The work at Virginia Commonwealth University is supported by the US National Science Foundation under grant 
CMMI-1301013.

\vfill

\pagebreak

\section{SUPPLEMENTARY MATERIAL}

Numerous theoretical analyses and characterization results are presented in ref. [16] of the main paper and the 
associated supplementary material which are not repeated here. They are to be viewed in conjunction with the material here.

We provide here energy-dispersive x-ray
spectra of nanowire spin-valve 
samples used in the experiments [Fig. 5], as well as room-temperature magnetization  (M-H curves) results of 
Co and Ni electrodeposited in nanoporous alimina hosts for 30 seconds and 4 minutes, respectively [Figs. 6 and 7].  
 The M-H curves were obtained with a Quantum Design Vibrating Sample Magnetometer.

\begin{figure}
\centering
\includegraphics[width=6in]{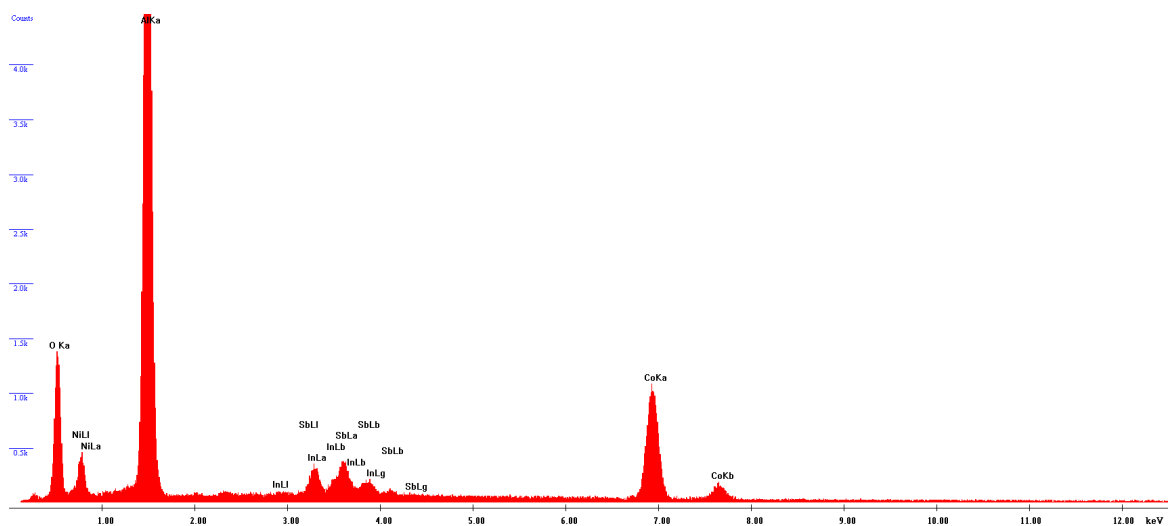}
\caption{Energy dispersive x-ray spectra of Co-InSb-Ni nanowires electrodeposited within 50-nm pores of anodic 
alumina films.}
\end{figure}


\begin{figure}
\centering
\includegraphics[width=6in]{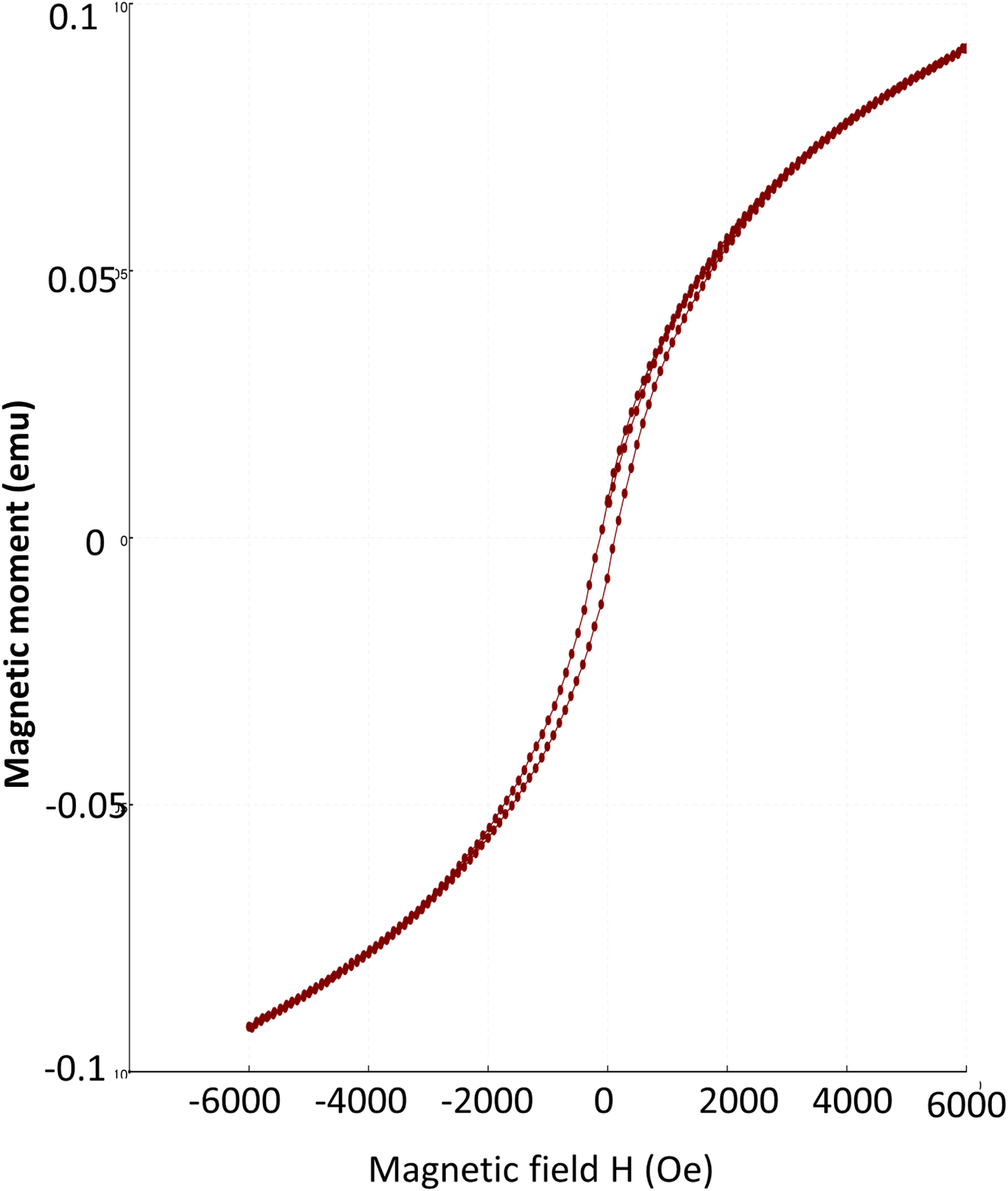}
\caption{Magnetization (magnetic moment) versus magnetic field characteristics of Co nanocontacts electrodeposited for 30 seconds 
in nanoporus anodic alumina film hosts. The measurements were made at room temperature and the magnetic field is directed parallel to
the axis of the pores into which Co is electrodeposited.} 
\end{figure}

\begin{figure}
\centering
\includegraphics[width=6in]{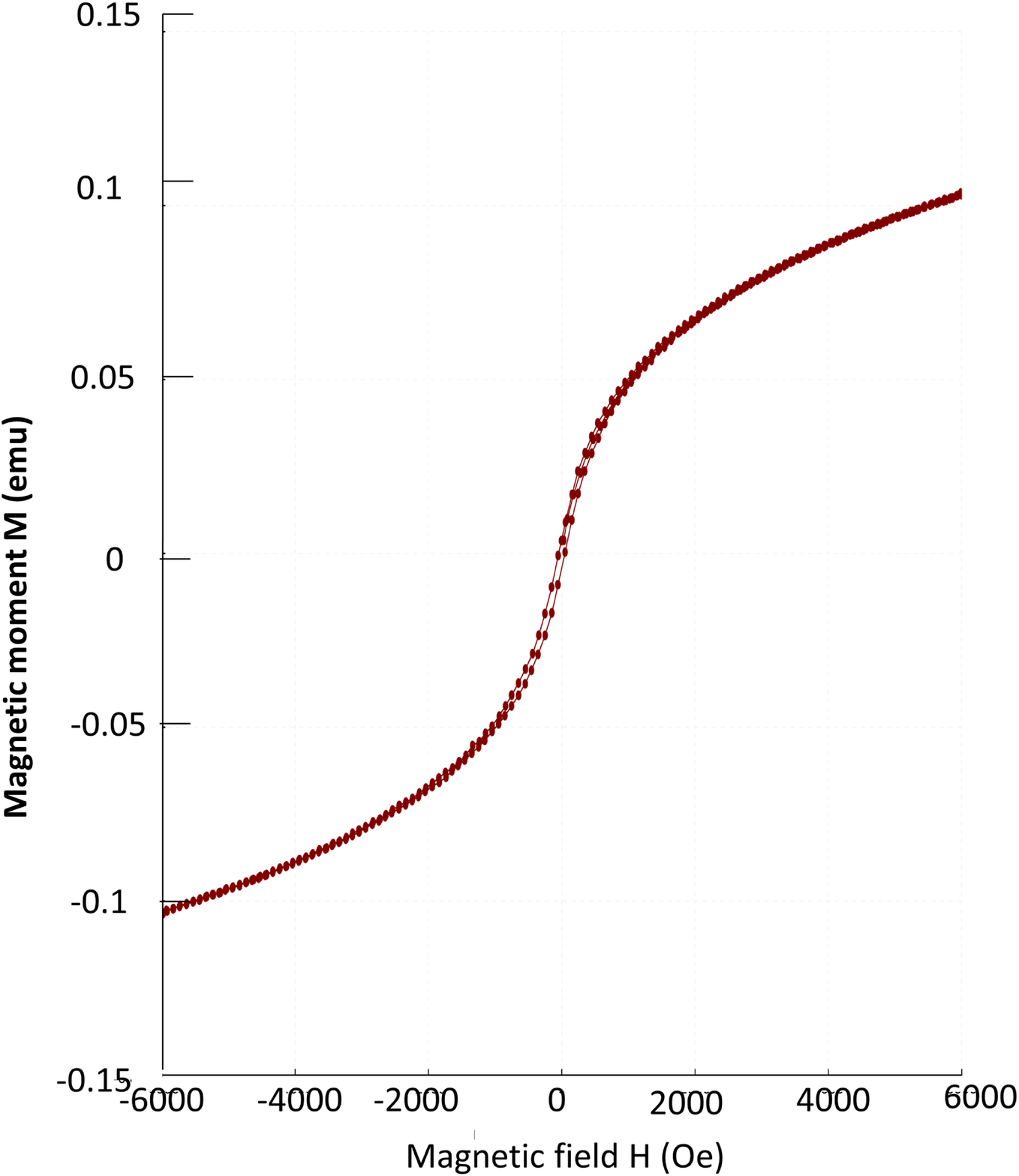}
\caption{Magnetization (magnetic moment) versus magnetic field characteristics
of 
Ni contacts electrodeposited for 4 minutes. 
The measurements were carried out at room temperature and the 
magnetic field is directed parallel to the axis of the pores.}
\end{figure}

\section{Transport model}
The resonant tunneling through defect sites that results in inversion of the spin valve resistance peak 
[discussed in the main paper and ref. [28]] does not require electrons to tunnel resonantly through the {\it entire} spacer layer; 
instead it requires resonant tunneling 
through one or more defect sites (approximated as ``point defects'') that are 
much smaller in physical extent than the spacer layer. 
If we draw the conduction band profile of the spacer layer, it will look like the diagram in Fig. 8. As long as there is 
a point defect close to a contact and electrons resonantly tunnel through it, the sign of the contact's spin polarization 
will be inverted. If there is no such point defect then the sign will not be inverted. That is why one sample showed sign inversion and 
another did not.
\begin{figure}
\centering
\includegraphics[width=6.5in]{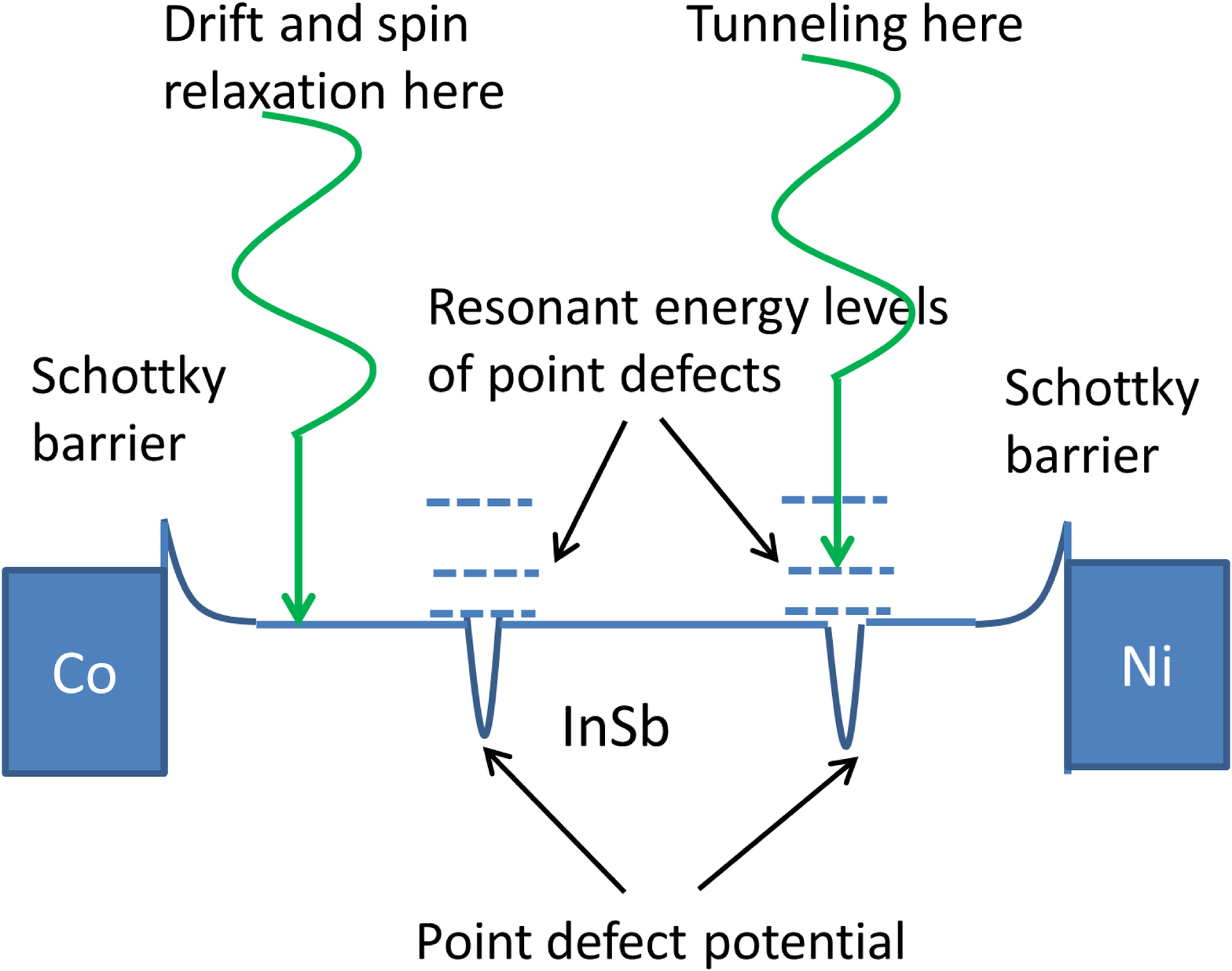}
\caption{Conduction band profile of the lowest subband in the spacer layer under no bias. The point defect potentials and 
their resonant 
energy levels are shown.
Transport between the defects occurs via drift, along with accompanying spin relaxation.} 
\end{figure}
Spin relaxation occurs during carrier drift between successive defect sites. Single subband occupancy reduces the
associated  relaxation rate 
by eliminating DP relaxation.

\section{The threshold magnetic field}
In the absence of any magnetic field, the Co contact's easy axis of magnetization is not along the pore axis, but perpendicular to it, as shown in the 
top panel of Fig. 9 because the 
thickness of the cobalt layer within the pore is smaller than the pore diameter of 50 nm. The Co
contact is roughly a cylinder of circular cross section whose cross-sectional diameter is 50 nm and height is much smaller than
50 nm. This happens because the Co layer is electrodeposited within 
the pore for a very short duration intentionally (our electrodepostion calibration predicts that the Co layer thickness will be around 20 nm). 
Hence, the Co layer's easy axis of magnetization will be in the plane perpendicular to the pore axis and the magnetization of the Co contact will be 
stable only in that plane. In other words, the Co contact's magnetization in the absence of an external magnetic field will be pointing perpendicular 
(or nearly perpendicular) to the pore axis as shown in the top panel of the figure below. 

A certain amount of external magnetic field will therefore be required to turn the magnetization of the Co contact by 90$^{\circ}$ and make it point 
along 
the pore axis so that it can inject/detect electrons whose spins are polarized along the pore axis. This magnetic field is the one required to overcome 
the Co 
contact's shape anisotropy energy barrier and make its magnetization point along the cylinder's axis as shown in the bottom panel of Fig. S5. 
This is the {\it threshold field} at which the magnetoresistance jumps because at or beyond this field, the Co contact preferentially
injects spins polarized along the pore axis, which are blocked by the Ni contact since its polarization is also along the pore axis 
but effectively inverted owing to resonant tunneling through the defect sites in the InSb spacer. 
\begin{figure}
\centering
\includegraphics[width=6.5in]{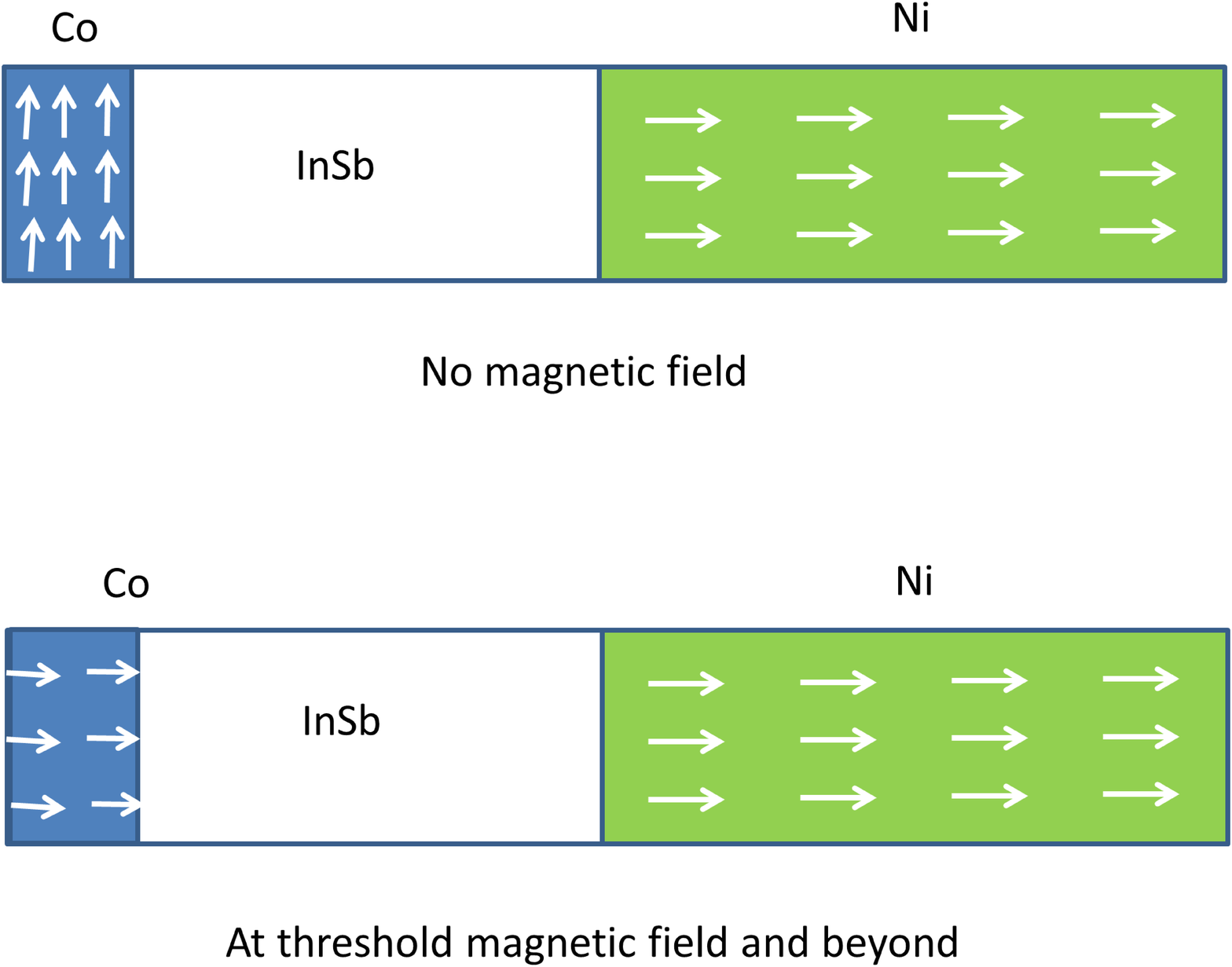}
\caption{The magnetizations of the two contacts below and above the threshold magnetic field.} 
\end{figure}

\end{document}